\documentclass{wqcd03}                 

\confname{QCD@Work 2003 - International Workshop on QCD, Conversano, Italy, 
14--18 June 2003}

\title{Quark determinant in domain-like gluon fields}

\author{A.C. Kalloniatis\addressmark{a}\thanks{Speaker at the Workshop.} 
and S.N. Nedelko\addressmark{b,c}
 }

\address[a]{Centre for the Subatomic Structure of Matter, 
University of Adelaide, Adelaide, Australia}
\address[b]{Institute for Theoretical Physics III, University of 
Erlangen-Nuremberg, Erlangen, Germany}
\address[c]{Bololiubov Laboratory of Theoretical Physics, JINR, Dubna, Russia}

\begin{document}

\begin{abstract}We address the computation of
the determinant for the Dirac operator
corresponding to a quark propagating in a background gluon field of the
following type: the gauge field is covariantly constant and self-dual 
inside a hypersphere and with quark fields satisfying
bag-like (chiral-violating) boundary conditions at the surface of
the hypersphere. We find that the parity odd part of the logarithm of the 
determinant corresponds  to
the Abelian anomaly. 
However the parity even part also
depends on the chiral angle associated with the boundary. This unexpected 
dependence is discussed.
\end{abstract}

\maketitle


\section{Introduction}

The profound physical phenomena expected to arise from
QCD, such as confinement, spontaneous chiral symmetry breaking
and consequences of the Abelian anomaly, have inspired a
range of simplified mathematical settings for the QCD equations
and searches for exact solutions to such problems. The exact
self-dual and finite action
solutions to the Yang-Mills classical equations known as the instanton
are one example of this. In the quark sector, a similar
endeavour has developed involving searches for exact solutions
to the Dirac eigenvalue spectrum for given nontrivial background
gluon fields, and in turn computations of the quark determinant
(and thus the quark free energy) in such gluonic backgrounds.
Often using zeta function regularisation or heat-kernel methods,
such problems have now been solved for various compact spaces
in various dimensions with and without
nontrivial gluon fields in the background. 
In this paper we shall address
such a computation for the quark determinant in what we
have previously called a ``domain-like'' gluon background
field, based on our earlier solution to the Dirac eigenvalue
problem for such a configuration. 

The elements of the problem are as follows: we consider a
four-dimensional Euclidean hypersphere of radius $R$. 
On the surface, fermion fields  satisfy
a chirality violating boundary condition
reminiscent of bag models of the nucleon, but now
for four-dimensional domains with nontrivial field
internal to the domain.  In various space-time dimensionalities,
recent works have addressed such boundary conditions
\cite{WD94,DW96,BK95,DAKB95,BFS99,Kir01,MaV03}
illustrating the relevance of the problem
to the realisation of chiral symmetries in QCD.
For example, in \cite{WD94} the fermion boundary condition is regarded as
a mechanism for dynamical generation of the CP-breaking $\theta$-term. 

An interesting feature of these boundary conditions is that they
render the spectrum of the Dirac operator
asymmetric under $\lambda\rightarrow -\lambda$ which in
turn has fascinating consequences. In particular, a careful analysis of
the fermion determinant leads to an additional asymmetry spectral
function \cite{DGS98,EBK98,BSW02}, as well as the usual zeta function type
contributions. In this respect, the four dimensional theory we tackle
resembles three dimensional field theories where parity
violating effects are well-known and accessed
through such asymmetries \cite{DGS98}.  The new element we introduce
here is the presence inside the hyperspherical region of
a non-vanishing covariantly constant, self-dual gluon field.

This type of gluon background field arises in 
 the ``domain model'' ansatz for the QCD vacuum
\cite{NK2001}, consisting
of an ensemble of such hyperspheres with randomly assigned
field orientations and self-duality or anti-self-duality. 
The fermion boundary condition in this case arises as a result
of requiring finite action fluctuations in the presence of
a singular field defining the boundary.
A chiral angle associated with fermion boundary condition here
becomes an additional random variable associated with a given
domain, such that averaging over the angle for an ensemble
of domains restores chiral symmetry for the ensemble \cite{NK2002}. 
In such a model, we have previously demonstrated confinement
of static quarks \cite{NK2001} and   have shown that the chirality of
low lying
Dirac modes exhibit strong correlation with the 
duality of the gluon field inside the
domain \cite{NK2002} reminiscent
of lattice QCD calculations \cite{Hor02,Xstudies,Gatt01}
and consistent with the dominance
of the chiral properties of the lowest modes in
condensate formation, as expressed in the well-known Banks-Casher relation
\cite{BaC80}. 
The computation of the quark determinant, 
tackled in this paper, is crucial for studying 
chiral symmetry realisation in the
domain model, and can be important for understanding this problem in QCD itself.
In particular, the determinant is precisely the relevant quantity
for obtaining the  Abelian anomaly in the path-integral
approach~\cite{Fuj80}. 
The  calculation  presented here recovers
the Abelian anomaly, which is identified as the parity odd part of the
logarithm of determinant. The anomaly gets contributions
from both the standard $\zeta$-function of the squared Dirac 
operator and the spectral asymmetry function, since in the
presence of chirality violating boundary conditions  
the spectral $\zeta$-function and 
spectral asymmetry function do not have direct physical meaning as 
parity even and odd respectively.
The dependence of the anomaly on the   
chiral angle turns out to be multivalued, which is to be  
expected since the topological charge here is not restricted to integer
values. The parity even part also shows a dependence
on the chiral angle (the parameter of boundary conditions),
which we consider to be more an artifact of the
incompletness of our calculation than an established
property of the determinant.  
Here we will only identify those potential contributions which are 
beyond the scope of the present calculation, and which can eliminate 
the chiral angle dependence of the parity even part.

\section{Boundary Conditions, Dirac operator and eigenvalues}
Our starting point is the eigenvalue problem for the
Dirac operator subject to the bag-like boundary condition
%
\begin{eqnarray}
\label{evp-0}
\!\not\!D\psi(x)&=&\lambda \psi(x),
\\
\label{quarkbc}
\!\not\!\eta(x)\gamma_5 e^{\vartheta \gamma_5}\psi(x)&=&\psi(x), \ x^2=R^2.
\end{eqnarray}
%
Here $\eta_\mu(x)=x_\mu/|x|$,  $D_\mu$ is 
the covariant derivative in 
the fundamental representation, 
%
\begin{eqnarray*}
D_\mu&=&\partial_\mu-i\hat B_\mu \\
&=&\partial_\mu + \frac{i}{2}\hat n B_{\mu\nu}x_\nu,
\\
\hat n&=&t^an^a, \\
\tilde B_{\mu\nu}&=&\pm  B_{\mu\nu},
\end{eqnarray*}
%
where the (anti-)self-dual tensor $B_{\mu\nu}$  is constant, and the 
Euclidean $\gamma-$matrices are in an anti-hermitean representation. 
To be specific, we will take the field as being self-dual.
The magnitude of the topological charge per domain will be
a useful quantity for later, and here it turns out to be \cite{NK2001}
just $q=B^2R^4/16$. 
As in \cite{WD94},
eigenvalues defined by Eqs.~(\ref{evp-0}) and (\ref{quarkbc})
are real for all real $\vartheta$.   
In this case the boundary condition  
for $\bar\psi$ required by the condition 
%
\begin{eqnarray}
\bar\psi(x)\!\not\!\eta(x)\psi(x)=0, \ x^2=R^2,
\nonumber
\end{eqnarray}
%
is given by hermitian conjugating Eq.~(\ref{evp-0}).
Complex values of the chiral angle require in general construction of a
bi-orthogonal basis.
Crucial to the solution derived in \cite{NK2002} was the decomposition
%
\begin{eqnarray}
\label{efr}
\psi=i\!\not\!\eta \chi + \varphi, \  \bar\psi=i \bar\chi \!\not\!\eta  + 
\bar\varphi,
\end{eqnarray}
%
$\varphi$ and $\chi$ having the same chirality. Solutions are then labelled
via the Casimirs and eigenvalues 
%
\begin{eqnarray*}
{\bf K}_1^2 & = & {\bf K}_2^2 \rightarrow 
            \frac{k}{2}(\frac{k}{2} + 1), \  k=0,1,\dots,\infty \\
K^z_{1,2} & \rightarrow & m_{1,2}, \\
 m_{1,2}&=&-k/2, -k/2+1,\dots,k/2-1,k/2, \\
k&=&0,1,2, \dots , 
\end{eqnarray*}
%
corresponding to the angular momentum operators
${\bf K}_{1,2} = \frac{1}{2} ({\bf L} \pm {\bf M})$
with ${\bf L}$ the usual three-dimensional angular momentum
operator and ${\bf M}$ the Euclidean version of the boost operator.
Thus the solutions for the self-dual background field can be written 
%
\begin{eqnarray}
\psi^{-\kappa}_{km_1}=i\!\not\!\eta\chi^{-\kappa}_{km_1}
+\varphi^{-\kappa}_{km_1}.
\label{psikappa}
\end{eqnarray}
%
The minus sign label on these solutions refers to
their negative chirality, $\gamma_5 \varphi=-\varphi, \ \gamma_5 \chi=-\chi$,
and is the only choice of chirality for which the boundary condition 
Eq.~(\ref{quarkbc})
can be implemented for the self-dual background field \cite{NK2002}. 
(For an anti-self-dual field, the chiralities become positive.)
The label $\kappa=\pm$ refers to the polarisation of the spinors
according to the combined colour-spin projection specified by
the projector
%
\begin{eqnarray}
O_{\kappa} = N_+ \Sigma_{\kappa} + N_- \Sigma_{-\kappa}
\end{eqnarray}
%
with 
%
\begin{eqnarray*}
N_{\pm} &=& \frac{1}{2}(1\pm \hat n/|\hat n|), \\
\Sigma_\pm  &=& \frac{1}{2}(1\pm{ {\bf\Sigma}\cdot{\bf  B}}/B)
\end{eqnarray*}
%
being respectively separate projectors for colour and spin polarizations.

In the $\zeta$-regularised determinant an arbitrary
scale $\mu$ appears and it is convenient to work with scaled variables
%
\begin{eqnarray}
\beta=\sqrt{2\hat B}/\mu, \ \rho=\mu R, \ \xi=\lambda(B)/\mu, \ 
\xi_0=\lambda(0)/\mu, \ 
\end{eqnarray}
%
and where the dimensionless quantity $z=BR^2/2=\beta^2\rho^2/4$
appears prominently. Then, 
as derived in \cite{NK2002}, 
the equations determining the (rescaled) eigenvalues $\xi$
for the two polarisations that appear in the problem
are
%
\begin{eqnarray}
\label{pmpm}
&&M(-\xi^2/\beta^2,k+1,z)
\\
&&+
e^{\vartheta}\frac{\xi\rho}{2(k+1)}M(1-\xi^2/\beta^2,k+2,z)=0
\nonumber
\end{eqnarray}
%
for $\kappa=-$
and
%
\begin{eqnarray}
\label{mppm}
&&M(k+2-\xi^2/\beta^2,k+2,z)
\\
&&+e^\vartheta\frac{\beta^2\rho}{2\xi}
\left[M(k+2-\xi^2/\beta^2,k+2,z)
\right.\nonumber \\
&&\left.-\frac{k+2-\xi^2/\beta^2}{k+2}
M(k+3-\xi^2/\beta^2,k+3,z)\right]=0
\nonumber
\end{eqnarray}
%
for $\kappa=+$. The confluent hypergeometric, or Kummer, functions
$M(a,b,z)$ emerged originally as solutions of the radial equation for
a scalar field in the presence of the above covariant constant
gauge field on a finite hypersphere \cite{NK2001}.

\section{Computing the determinant}
We now consider the quark determinant
in a single self-dual domain of  volume $v=\pi^2 R^4/2$ 
%
\begin{eqnarray}
{\rm det}_\vartheta\left(\frac{i\!\not \!D}{i\!\not\! \partial}\right)  
&=&\exp\sum_{k,n,m_1}\left(\ln\frac{\lambda^\kappa_{kn}(B)}{\mu}
-\ln\frac{\lambda^\kappa_{kn}(0) }{\mu}\right)
\nonumber\\
&=&\exp\left\{-\zeta'(s)\right\}_{s=0}
\label{det1}
\end{eqnarray}
%
where the arbitrary scale $\mu$ plays a role similar to
the renormalization point in other regularisation prescriptions.
The zeta function $\zeta(s)$ breaks up into
two parts \cite{DGS98}
respectively symmetric (S) and antisymmetric (AS) with respect to 
$\lambda\to-\lambda$   
%
\begin{eqnarray}
\zeta(s)=\zeta_{\rm S}(s)+\zeta_{\rm AS}(s),
\end{eqnarray}
%
with
%
\begin{eqnarray}
\zeta_{\rm S}(s)& =& \frac{1}{2}\left(1+e^{\mp i\pi s}\right)
\zeta_{\!\not\!D^2}(s/2),\nonumber  \\
\zeta_{\rm AS}(s)&=&\frac{1}{2}\left(1-e^{\mp i\pi s}\right)\eta(s)
\nonumber
\end{eqnarray}
%
and the key quantities devolve into the zeta function for the
squared Dirac operator and the asymmetry function respectively:
%
\begin{eqnarray}
\zeta_{\!\not\!D^2}(s)&=&\sum_{k,n,\kappa}(k+1)
\left(\frac{\mu^{2s}}{[\lambda^\kappa_{kn}(B)]^{2s}}
\right. \nonumber \\
&&\left. -
\frac{\mu^{2s}}{[\lambda^\kappa_{kn}(0)]^{2s}}
\right)
\label{zeta1}
\\
\eta(s)&=&\mu^s\sum_{k,n,\kappa}(k+1)\left(
\frac{{\rm sgn}(\lambda^\kappa_{kn}(B))}{|\lambda^\kappa_{kn}(B)|^{s}}
\right.\nonumber \\
&&\left.
-\frac{{\rm sgn}(\lambda^\kappa_{kn}(0))}{|\lambda^\kappa_{kn}(0)|^{s}}
\right).
\label{eta1}
\end{eqnarray}
It should be stressed  that
in the presence of bag-like boundary conditions
$\zeta_{\rm S}$ and $\zeta_{\rm AS}$ do not have the meaning of 
parity conserving and parity violating terms
since a parity transformation in terms of eigenvalues is given by 
$\lambda(\vartheta) \to -\lambda(-\vartheta)$
and both spectral functions contain parity conserving and violating terms.
Thus the determinant for a given parameter $\vartheta$ is defined by
%
\begin{eqnarray}
\zeta'(0)=\left(\frac{1}{2}\zeta'_{\!\not\!D^2}(0)
\pm i\frac{\pi}{2}\zeta_{\!\not\!D^2}(0)\mp i\frac{\pi}{2}\eta(0)\right),
\label{zetaprime}
\end{eqnarray}
%
with the normalization chosen in Eq.(\ref{det1}) such that $\zeta'(0)$ vanishes
as $B\rightarrow 0$.

Spectral sums can be computed using a representation of the
sum as a contour integral of the logarithmic derivative
of the function whose zeroes determine the spectrum, 
%
\begin{equation}
\sum_{\lambda} {1 \over {\lambda^s}} = \frac{1}{2\pi i}
\oint_{\Gamma} \frac{d\xi}{\xi^s} \frac{d}{d\xi}\ln f(\xi),
\label{intrep}
\end{equation}
%
see for example~\cite{BSW02},
where the zeroes of $f(\xi)=0$ are $\xi=\lambda$ and the contour is
chosen such that all zeroes are enclosed. With real parameter
$\vartheta$, the poles lie on the real
axis, and there is no pole at the origin for any $\vartheta$. 
By deforming the contour and accounting for the vanishing of
contributions to the integral at infinity, the expressions arising from Eq.~({\ref{intrep})
can be transformed into real integrals. The following representations
for the two spectral functions are eventually obtained
%
\begin{eqnarray}
\zeta_{\not D^2}(s)&=&\rho^{2s}\frac{\sin(\pi s)}{\pi}
\sum_{k=1}^\infty k^{1-2s} \nonumber \\
&&\times 
\int_0^\infty \frac{dt}{t^{2s}}\frac{d}{dt}\Psi(k,t|z,\vartheta),
\\
\eta(s)&=& \rho^{s}\frac{\cos(\pi s/2)}{i\pi}
\sum_{k=1}^\infty k^{1-s} \nonumber \\
&&\times
\int_0^\infty \frac{dt}{t^{s}}\frac{d}{dt}\Phi(k,t|z,\vartheta),
\end{eqnarray}
%
with $\Psi$ and $\Phi$ being the sum of contributions from the
two polarisations, taking the form
%
\begin{eqnarray}
\Psi(k,t|z,\vartheta) &= & \sum_{\kappa=\pm}\ln \left(
{ {A_{\kappa}^2(k,t|z)+e^{2\vartheta} B_{\kappa}^2(k,t|z)}} \over
{A^2(k,t) + e^{2\vartheta} B^2(k,t)} \right)
\nonumber \\
\Phi(k,t|z,\vartheta) &=& \sum_{\kappa=\pm}\ln \left(
{ {A_{\kappa}(k,t|z)+i e^{\vartheta} B_{\kappa}(k,t|z)} \over
{A_{\kappa}(k,t|z) -i e^{\vartheta} B_{\kappa}(k,t|z)}} \right)
\nonumber 
\end{eqnarray}
%
and where 
%
\begin{eqnarray}
\label{AB}
A_{-}(k,t|z)&=&M(\frac{k^2t^2\rho^2}{4z},k+1,z), 
 \nonumber\\
B_{-}(k,t|z)&=&\frac{kt\rho}{2(k+1)}
M(1+\frac{k^2t^2\rho^2}{4z},k+2,z),
\nonumber\\
A_{+}(k,t|z)&=&M(-\frac{k^2t^2\rho^2}{4z},k+1,-z), 
\nonumber\\
B_{+}(k,t|z)&=&\frac{2z}{kt\rho}
\left[M(-\frac{k^2t^2\rho^2}{4z},k+1,-z)
\right. \nonumber \\
&&-\frac{k+1+\frac{k^2t^2\rho^2}{4z}}{k+1}\nonumber \\
&&\left. \times 
M(-\frac{k^2t^2\rho^2}{4z},k+2,-z)\right],
\nonumber\\
A(k,t)&=&\frac{2^{k}k!}{(kt\rho)^{k}}I_k(kt\rho), \nonumber \\
B(k,t)&=&\frac{2^{k}k!}{(kt\rho)^{k}}I_{k+1}(kt\rho).
\end{eqnarray} 
%
The Bessel functions $I_k$ emerge from the limit $B\rightarrow 0$
with the normalization of the determinant as specified above.
Analytical continuation of $\zeta_{\not D^2}(s)$  and $\eta(s)$ 
can be done by means of standard methods 
(for an extensive review see~\cite{Elizbook}).
These functions can be
related to functions such as the Riemann
zeta function $\zeta_R(s)$ with known analytical properties.
Our particular example of such a calculation is clearly
technically quite complicated.
First of all, expansions of the confluent hypergeometric functions
in $1/k$, similar to the Debye expansion of Bessel functions, is required,
%
\begin{eqnarray}
M(\frac{k^2t^2\rho^2}{4z},k+1,z) &=& 
C(t\rho,k)\sum_{n=0}^\infty {{M_n(t\rho,z)}\over k^n}.
\label{Mexpand}
\end{eqnarray}
%
The form of the prefactors and the $M_n(x,z)$
functions will be presented in detail elsewhere \cite{NKprep}.
It suffices to say that 
the prefactor $C(x,k)$ cancels out of $\Phi$ and $\Psi$
after which the procedure is straightforward if somewhat tedious:
the expansions Eq.~({\ref{Mexpand}) are inserted in $\zeta(s)$
and $\eta(s)$
so that the integrals over $t$ can be evaluated term by term. After 
exchange of order of summation over $k$  and $n$ 
(which is the most subtle step here), 
sums over $k$ can be read off in terms of the 
Riemann zeta function.  
The resulting $\zeta_{\not D^2}(s)$ has the structure
%
\begin{eqnarray}
\zeta_{\not D^2}(s)=s \rho^{2s} {\sin(\pi s)\over i\pi}
\sum_{n=0}^\infty \zeta_{\rm R}(2s+n-1)f(z^2|n) 
\nonumber\\
+\delta \zeta_{\not D^2}(s),
\label{struct}
\end{eqnarray} 
%
where the term in the second line denotes 
those contributions coming from 
interchange of the order of summations over $n$ and $k$,
which are potentially present.
A similar structure appears for the asymmetry spectral function as well.
Finally we are interested in the decomposition of this expression around $s=0$.
The result for the first line in Eq.(\ref{struct}) comes from the 
$n=2$ term alone and can be calculated with relative ease 
since only the lowest coefficients $M_1$ and $M_2$
in Eq.~(\ref{Mexpand}) contribute.  However, as  occurs
in even simple problems \cite{Elizbook}, 
the second term in Eq.(\ref{struct}) is much more difficult to
compute. To achieve this one needs to know coefficients $f(z^2|n)$ as a 
function of  continuous variable $n$. 
Below we will concentrate on the contribution
given in the first line alone, bearing in mind 
the necessity of a complete analysis. 
The final results for the first term in Eq.~(\ref{struct})
and its analogue in the asymmetry function $\eta(s)$
are then summarised in the following equations:
%
\begin{eqnarray}
\zeta_{\not D^2}(0)&=&0, \nonumber \\
\zeta'_{\not D^2}(0)&=&z^2\left[
-\frac{1}{4}-\ln2+\ln\left(1+e^{2\vartheta}\right)
\right]
\nonumber\\
\eta(0)&=&\frac{z^2}{2}+\frac{z^2}{\pi}{\rm Arctan}({\rm sinh}(\vartheta))
\end{eqnarray} 
%
and hence
%
\begin{eqnarray}
\label{finalzeta}
\zeta'(0)&=& \frac{z^2}{2}\left[-\frac{1}{4}-\ln2\pm 
i\frac{\pi}{2}\right.
 \\
&&\left.+\ln\left(1+e^{2\vartheta}\right)
\mp i {\rm Arctan}({\sinh}(\vartheta))\right]+\delta \zeta'(0),
\nonumber
\end{eqnarray} 
%
where $\delta \zeta'(0)$  denotes contributions potentially coming from the 
above-discussed non-commutativity of summations, which have been not 
calculated here. For an anti-self-dual
background field, we merely have to perform a parity transformation
on the fields. For the spectrum it involves 
$\lambda(\vartheta)\rightarrow -\lambda(-\vartheta)$. For the
determinant, one need only take $\vartheta\rightarrow-\vartheta$
and $\eta(s)\rightarrow -\eta(s)$ independently. 

\section{Discussion and Conclusions}
For interpretation of the result Eq.~({\ref{finalzeta})
one needs to continue $\vartheta$ to imaginary values. 
The appropriate
continuation is $\vartheta\rightarrow i \alpha+i\pi/2$.
In this case, one can cleanly separate parity odd and even parts since for
the self-dual field one gets

\begin{eqnarray}
\zeta'(0)\to\frac{z^2}{2}\left[-\frac{1}{4}+\ln(1\pm\cos(\alpha))
+i (\alpha + \pi \nu)\right],
 \nonumber
\end{eqnarray}
 
with $\nu \in Z$,
while the anti-self-dual field leads to

\begin{eqnarray}
\zeta'(0)\to\frac{z^2}{2}\left[-\frac{1}{4}+\ln(1\pm\cos(\alpha))
-i (\alpha + \pi \nu)\right].
 \nonumber
\end{eqnarray} 
The imaginary part is given in a form which manifests
the $2\pi$ periodicity in $\alpha$.
Recalling that the absolute value of 
the topological charge associated with the gluon field in a domain
is $q=z^2/4$, the parity odd part
becomes 
\begin{eqnarray}
\pm 2i q (\alpha \ {\rm mod}\ \pi )
\nonumber
\end{eqnarray}
with the sign correlated with the duality of the field,
and can be regarded as the axial anomaly.
This agrees with the estimation of \cite{WD94}
obtained in a more general approach, up to a term $\pm iq \pi$.
As far as we can see this difference occurs 
due to contributions in our calculation from both $\zeta_{\not D^2}'(0)$
and the asymmetry spectral function $\eta(0)$, which provides
an extra $\pm iq \pi$ to give the above anomaly term. 
The contribution of the asymmetry spectral function was not considered in
\cite{WD94}.

Final conclusions about the validity of the chiral angle dependence
of the parity even part, as well as an analysis of the strong 
background field limit, necessarily require computation  
of the potentially nonzero term $\delta\zeta_{\not D}$ in Eq.(\ref{struct}).

\section*{Acknowledgments}
S.N.N. was supported by the DFG under contract SM70/1-1 and, 
partially, by  BFBR grant 01-02-1720. 
 We acknowledge the hospitality of the Institute
for Theoretical Physics III, University of Erlangen-Nuremberg
where substantial parts of this work was done.
A.C.K. is supported by a grant of the Australian Research Council.

\end{document}